\documentclass[useAMS,usenatbib]{mn2e}
\usepackage{graphicx}
\def\msol{M$_{\odot}$}

\def\Ne2{\mbox{[Ne\,{\sc ii}]}}

\def\arcsec{\,''}

\title[Nuclear embedded star clusters in NGC~7582]
{Nuclear embedded star clusters in NGC 7582\thanks{Based on observations
obtained with the ESO Very Large Telescope}}
\author[M. Wold and E. Galliano]
       {M. Wold$^{1}$\thanks{E-mail:mwold@eso.org} and E. Galliano$^{2}$ \\
$^{1}$European Southern Observatory, Karl-Schwarzschild str.~2,
              85748 Garching bei M{\"u}nchen, Germany \\
$^{2}$European Southern Observatory, Casilla 19001, Santiago 19, Chile}

\begin{document}

\date{Accepted 1988 December 15. Received 1988 December 14; 
in original form 1988 October 11}

\pagerange{\pageref{firstpage}--\pageref{lastpage}} \pubyear{2006}

\maketitle

\label{firstpage}

\begin{abstract}
We report on the discovery of several compact regions of mid-infrared emission
in the starforming circum nuclear disk of the starburst/Seyfert~2 galaxy 
NGC~7582. The compact sources do not have counterparts in the optical and 
near-infrared, suggesting that they are deeply embedded in dust. We use the
\mbox{[Ne\,{\sc ii}]}12.8$\umu$m line emission to estimate the emission measure
of the ionized gas, which in turn is used to assess the number of ionizing 
photons. Two of the brighter sources are found to have ionizing fluxes of 
$\sim 2.5\times 10^{52}$, whereas the fainter ones have 
$\sim 1\times10^{52}$ photons\, s$^{-1}$. Comparing with a one Myr old 
starburst, we derive stellar masses in the range 
3--5$\times$10$^{5}$ M$_{\odot}$ and find that the number of O-stars in each 
compact source is typically 0.6--1.6$\times$10$^{3}$. 
We conclude that the compact mid-infrared sources are likely to be 
young, embedded star clusters, of which only a few are known so far. 
Our observation highlights the need for high resolution mid-infrared imaging
to discover and study embedded star clusters in the proximity of active 
galactic nuclei. 
\end{abstract}


\section{Introduction}

Starburst activity in nearby galaxies has, with the superior resolution of the
Hubble Space Telescope (HST), been resolved into discrete massive young 
clusters since the pioneering work by \citet{Holtzman92}. In the literature, 
these clusters are usually 
referred to as super star clusters (SSC) or young massive clusters (YMCs). 
Their masses are typically $\geq 10^{5}$ \msol, radii $\le 5$ pc and ages 
$\le$ 100 Myr. Most known young clusters in the Milky Way have masses below
$\sim 10^{4}$ \msol~\citep{Figer99}, but there also exist examples of YMCs
in the Milky Way, such as the Westerlund~1 cluster which has a mass of
$\sim 10^{5}$ \msol~\citep{clark05}.   
The most nearby analog to a YMC is the 30 Dor cluster in the Large Magellanic 
Cloud.  

YMCs are blue compact sources, whose light is dominated by massive stars. As 
in the case of massive star association formation, the progenitors of YMCs are
thermal radio sources deeply embedded in dust, e.g.\ the ultra dense 
\mbox{H\,{\sc ii}}~regions, first discovered in Henize\,2-10 by 
\citet{Kobulnicky99}. These objects 
are sources of mid-infrared (MIR) continuum and bright nebular line emission.
Because of high dust extinction, they are usually very faint in the visible, 
and even in the near-infrared (NIR). The first evidence for the existence of 
such MIR sources was unveiled by the Infrared Space Observatory (ISO) in the 
Antennae Galaxies NGC4038/9 \citep{Mirabel98}. Most of the MIR flux
from this galaxy complex comes from a small, optically insignificant, region. 
Subsequent NIR spectroscopy by \citet{Gilbert00} suggests that the related 
cluster formed about 4 Myr ago, has a mass of $\approx$16$\times$$10^6$ 
\msol, contains massive stars with $T_{\rm eff}$ up to $3.9\times10^4$\,K, and
is deeply embedded in dust, with $A_v \sim 10$.
  
Embedded YMCs have so far been observed in only a few galaxies: the Antennae 
and Henize\,2-10, as already mentioned, NGC 5253 \citep*{Gorjian01}, SBS 
0335-052 \citep{Plante02}, IIZw40 \citep{Beck02} and NGC1808 and NGC1365 
\citep{Galliano05a}.

Here we report on the discovery of bright, compact MIR sources in the 
composite starburst/Seyfert2 galaxy NGC~7582, using a VLT/VISIR image centered
on the \mbox{[Ne\,{\sc ii}]}12.8$\umu$m emission line. We present the data 
in terms of suggestive evidence that the MIR sources are embedded YMCs, and 
use the \mbox{[Ne\,{\sc ii}]} luminosity to estimate physical properties, 
such as the number of ionizing photons and the number of massive stars within 
each source. We also emphasize that MIR imaging at high resolution is 
mandatory to study embedded star formation in the vicinity of active galactic
nuclei (AGN). 

\section{The data}
\label{section:data}

The data were obtained at the VLT-Melipal telescope with VISIR, a combined 
imager and spectrograph for observations in the $N$ and $Q$ bands 
(corresponding to the two atmospheric windows at 8--13 and 16.5--24.5$\umu$m, 
respectively). A high-resolution spectrum and one narrow-band image 
centered on the \mbox{[Ne\,{\sc ii}]}12.8$\umu$m line were obtained during the
period September 29 to October 2, 2004. The VISIR detector is a SiAs 
256$\times$256 DRS array, and a pixel scale of 0\farcs127 was used, yielding a
field of view of 32\farcs5$\times$32\farcs5. 
Total integration time was one hour for both image and spectrum, and  
conditions on both nights were clear with humidity below 10\%.
The spectrum was taken in the high-resolution 
\mbox{[Ne\,{\sc ii}]}12.8$\umu$m
Echelle mode using a 0\farcs75 wide slit, giving a spectral resolution
$R\approx16000$, or $\approx$22 km\,s$^{-1}$.
The diffraction limit is reached at 12.8$\umu$m, hence both the spectrum
and the image have the optimal spatial resolution of 0\farcs38.
More details regarding the data are presented by 
Wold et al.\ (in preparation). 

NGC~7582 has a redshift of $z=0.00525$, hence one arcsec corresponds to 
$\approx 108$ pc in the galaxy rest 
frame\footnote{We assume $H_{0}=70$ km\,s$^{-1}$\,Mpc$^{-1}$}, and the
\mbox{[Ne\,{\sc ii}]} line is redshifted to 12.877$\umu$m. The 
\mbox{[Ne\,{\sc ii}]} 
narrow-band filter has $\lambda_{\rm c} = 12.81\umu$m and 
half band width 0.21$\umu$m, thus encompassing \mbox{[Ne\,{\sc ii}]} emission 
from the galaxy.

\begin{figure*}
\centering
\includegraphics[scale=0.45]{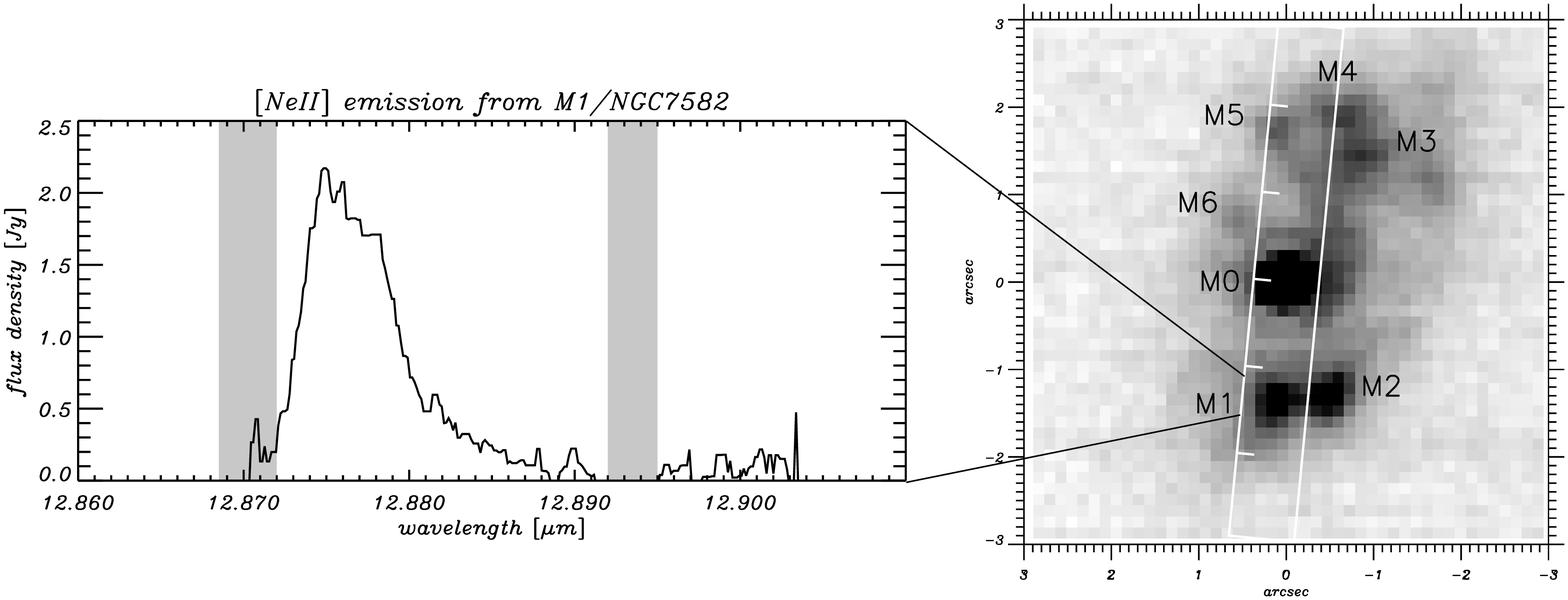}
\caption{Narrow-band image and spectrum centered on the \Ne2 line.
The spectrum has not been continuum subtracted.
The slit position is marked on the image. North is up and East is to the left.}
\label{figure:data}
\end{figure*}

The \Ne2 narrow-band image displayed in Fig.~\ref{figure:data} shows
the central 6\arcsec$\times$6\arcsec\, around the AGN.
The MIR emission is resolved into several components, labelled M0--M6, 
where M0 corresponds to the AGN. There are two noticeably bright, dense knots 
of \Ne2 emission to the South, labeled M1 and M2. 
The central AGN is unresolved whereas the full width at half 
maximum (FWHM) of M1 and M2 are 0\farcs60 and 0\farcs65, respectively. M2 
appears slightly extended in the SE-NW direction. 
The compact sources 
are themselves embedded in a larger and more diffuse envelope roughly 300--400
pc across. 
The position of the slit is 
indicated in the image, and the high-resolution spectrum of the \Ne2
line from M1 is shown to the left. 

The measured fluxes and flux densities of the compact sources are listed 
in Table~\ref{table:photometry}. The aperture diameters used correspond to 
1$\times$ and 2$\times$ the diffraction-limited resolution.
We estimated \Ne2 line fluxes from the 
narrow-band image by calibrating the image using the M1 line flux from the 
spectrum. The accuracy of the line fluxes derived from the narrow-band image 
in this manner is approximately 10~\%. The spectrum shows that there is no
continuum detected from M1. The noise level is $\approx0.09$ Jy, so the 
3$\sigma$ upper limit on the continuum is 0.27 Jy. By calibrating the 
narrow-band image using the M1 spectrum, we assume that the other 
sources M2--M6 also have non-detectable continua. 

\begin{table}
\caption{Measured quantities.
Columns (2) and (3) show flux density within 
aperture diameters 0\farcs38 and 0\farcs76, respectively. Columns (4) and (5)
show line flux, also within apertures of 0\farcs38 
and 0\farcs76.}
\label{table:photometry}
\begin{tabular}{lrrrr}
\hline
Source & \multicolumn{2}{c}{\Ne2 flux density} & \multicolumn{2}{c}{\Ne2 flux} \\ 
       & \multicolumn{2}{c}{(mJy)} & \multicolumn{2}{c}{(10$^{-16}$ W\,m$^{-2}$)}\\
 (1)   & (2) & (3) & (4) & (5) \\
\hline
M1 &      45.3$\pm$4.5 &  131.2$\pm$7.7 & 1.5$\pm$0.1$^{*}$ & 2.5$\pm$0.1$^{*}$ \\
M1 &      45.3$\pm$4.5 &  131.2$\pm$7.7 & 1.50$\pm$0.15 & 4.35$\pm$0.26 \\
M2 &      43.9$\pm$4.5 &  126.7$\pm$7.6 & 1.46$\pm$0.15 & 4.20$\pm$0.25 \\
M3 &      29.4$\pm$3.6 &  103.2$\pm$6.8 & 0.98$\pm$0.12 & 3.42$\pm$0.23 \\
M4 &      29.0$\pm$3.6 &   95.9$\pm$6.6 & 0.96$\pm$0.12 & 3.18$\pm$0.22 \\
M5 &      23.1$\pm$3.2 &   68.3$\pm$5.6 & 0.77$\pm$0.11 & 2.27$\pm$0.18 \\
M6 &      17.6$\pm$2.8 &   59.7$\pm$5.2 & 0.59$\pm$0.09 & 1.98$\pm$0.17 \\
Envelope$^{**}$ &  2334.5$\pm$32.5 & ... & 77.39$\pm$1.08 & ... \\
\hline
\end{tabular}
\begin{list}{}{}
\item[$^{*}$] Measured from the spectrum
\item[$^{**}$] Aperture diameter 7\farcs62
\end{list}
\end{table}

\begin{figure*}
\centering
\includegraphics[scale=0.8]{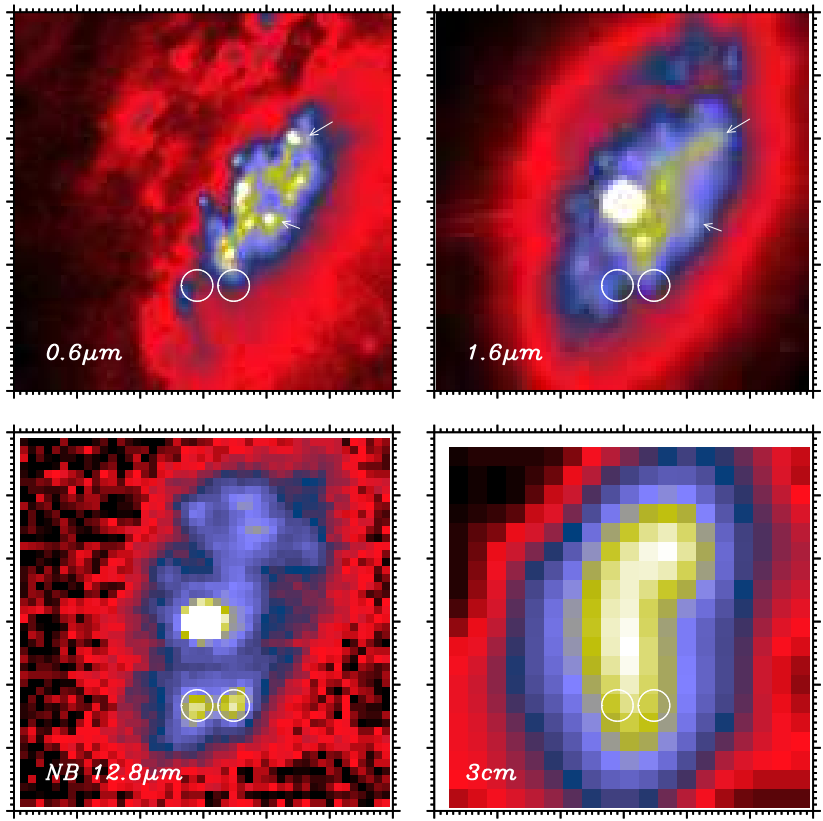}
\caption{Comparison with data at other wavelengths. Top row, from left
to right: HST/WFPC2 $F606W$ and HST/NICMOS $F160W$.
Bottom row, from left to right: 
the VLT/VISIR narrow-band image and a 3cm ATCA radio map. The two 
circles South of the nucleus 
indicate the location of the two MIR sources, M1 and M2. The two small
arrows in the HST images point to the stars that were used for 
aligning. Each major tick mark corresponds to 1\arcsec.
North is up, and East is to the left.}
\label{figure:img_composition}
\end{figure*}


\section{Comparison with data at other wavelengths}

\label{section:nbimg}

NGC~7582 is known to have a circum nuclear kpc-scale disk with active star 
formation, containing significant amounts of dust (\citealt{morris85,
regan_mulchaey99,sosabrito01}). This is however the first time that the disk 
has been viewed at high resolution in the MIR, and the nuclear starburst 
is resolved into several compact components embedded in a larger, more 
diffuse envelope. The compact sources, in particular M1 and M2, are 
reminiscent of young, massive, embedded star clusters. We searched the 
literature and various data archives in order to find counterparts to the
MIR sources at other wavelengths, but no clear identifications can be made.
A selection of data from optical to radio is shown in 
Fig.~\ref{figure:img_composition}. 

The top row of Fig.~\ref{figure:img_composition} shows two HST images, taken 
with WFPC2 and NICMOS through the $F606W$ and $F160W$ filters, respectively 
(programmes 8597 and 7330). Both HST images are dominated by stellar 
continuum, and the expected locations of M1 and M2 are marked with white open 
circles. No clear counterparts to M1 and M2 can be seen, and there is also no
clear counterparts to the other fainter sources M3--M6.

Although no counterparts are found in continuum optical and NIR, there is 
extended emission from the starburst in narrow-band images centered
on \mbox{[Fe\,{\sc ii}]}1.64$\umu$m and Br$\gamma$ at 2.165$\umu$m 
(\citealt{sosabrito01}, fig.~3).
\mbox{[Fe\,{\sc ii}]} traces shocks excited by supernova remnants 
\citep{vr97}, and indeed \citet*{reunanen03} observe a high 
\mbox{[Fe\,{\sc ii}]}/Br$\gamma$ intensity 
ratio typical of shock excitation. This might indicate that the most massive 
stars in the starburst have already exploded as supernovae. 
The resolution in the \mbox{[Fe\,{\sc ii}]} and Br$\gamma$
narrow-band images is however not as good as in the VISIR image, making it 
difficult to tell where the \mbox{[Fe\,\sc{ii}]} and Br$\gamma$
emission comes from exactly. Some of it may come from 
the compact sources seen in the VISIR image, but it is also possible that a 
large contribution comes from the diffuse envelope.  

The M2 source is clearly seen in the deconvolved 11.9$\umu$m image by 
\citet*{siebenmorgen04}, but M1 is either weak, or undetected. This morphology,
with a weak or undetected M1 and a brighter M2 source, is also recognized
in the narrow-band H$_{2}$ (2.12$\umu$m) image by \citet{sosabrito01}.
A host-galaxy subtracted $L$-band image by \citet*{prieto02} unveils part of 
the circum nuclear starburst disk, probably due to PAH emission at 3.3$\umu$m. 

The bottom row in Fig.~\ref{figure:img_composition} shows the VISIR
image and an ATCA radio map obtained at 8.9 GHz 
\citep{morganti99}. Young, dust-embedded star clusters are expected to 
show thermal free-free emission at radio wavelengths, but 
the resolution in the ATCA map is $\approx$1\arcsec, not sufficient to 
match the resolution of the VISIR image. Extended emission from the starburst 
can clearly be seen, but part of the radio emission may also be non-thermal 
synchrotron emission from the AGN. When the 
Atacama Large Millimeter Array (ALMA) becomes available, it will be possible
to separate the different components at mm and sub-mm wavelengths 
and study the starburst at great detail. 


\section{Physical parameters of the MIR sources}

\Ne2 is the most common Ne ion in H{\sc ii} regions, ionized by a broad
range of stellar types. It is collisionally excited, and the line at 
12.8$\umu$m results from spontaneous radiative deexcitation. 
In typical star clusters, the electron density, $n_{\rm e}$, is a factor of 
10 or more lower than the critical density for  
collisional thermalization of the 12.8$\umu$m transition, 
$n_{c}=5.4\times10^{5}$ cm$^{-3}$ \citep{johnson_kobulnicky03}. Because
$n_{\rm e} \ll n_{\rm c}$, the emission measure of the
ionized gas can be estimated from the \Ne2 line flux, which 
in turn can be used to predict the thermal free-free radio emission, the number
of ionizing photons and the number of massive stars \citep{Keto99}. 
We proceed along the lines of the analysis 
done by \citet{Keto99}, and calculate the emission measure by using that
the \Ne2 line intensity depends on the number of atoms in the upper state:
\begin{equation}
I =3\times 10^{-8}\left(\frac{10^{4} {\rm K}}{T_{\rm e}}\right)^{\frac{1}{2}} 
\exp(-\frac{h\nu}{kT_{\rm e}})\int 
\frac{n_{\rm e}}{(n_{\rm e} + n_{\rm c})} \gamma_{\rm Ne} n_{\rm e}^{2} 
{\rm d}l
\end{equation}
(\citealt{Keto99} and references therein\footnote{\rm The constant 
3$\times$10$^{8}$ in Keto et al.'s eq.~2 is probably a typing error, and should
be 3$\times$10$^{-8}$ as used here.}). 
The line intensity, $I$, has units of 
erg\,s$^{-1}$\,cm$^{-2}$\,sr$^{-1}$ and the emission measure, 
$EM = n_{\rm e}^{2}{\rm d}l$, has units of cm$^{-6}$\,pc. We assume electron 
temperature and density of $T_{e}=10^{4}$ K and $n_{e} \ll n_{c}$, 
respectively, and that the abundance of Ne relative to H is 
$\gamma_{\rm Ne} = 0.83$ \citep{Keto99}. From the flux measurements in 
Table~\ref{table:photometry}, we derive \Ne2 intensities by dividing 
with the solid angle, $\Omega$, subtended by the source, taken to be 
equal to the area of the circular aperture used for photometry. 
Table~\ref{table:emtab} lists the derived emission measures.

The emission measure is directly related to the optical depth, 
$\tau^{\rm ff}_{\nu}$, in the free-free continuum (\citealt{Keto99}, eq.~4 ), 
and we estimate optical depths at 5 GHz of $\approx$0.01--0.02. The optically
thin approximation for brightness temperature, 
$T_{\rm b} = T_{\rm e}\tau^{\rm ff}_{\nu}$, is therefore valid. We proceed with
calculating the radio flux density at 5 GHz arising from free-free processes 
from the brightness temperature
given that $S_{\nu} = 2\nu^{2}kT_{\rm b} / c^{2}\Omega$, see third column
of Table~\ref{table:emtab}.

\begin{table*}
\caption{Derived quantities of the sources. 
The size of the envelope is taken to be equal to the area of 
a circle with 7\farcs62 diameter.}
\label{table:emtab}
\begin{tabular}{lrrrrrrrr}
\hline
Source & $EM$ & $S^{\rm ff}_{5 GHz}$ &  
         $\log N_{\rm ph}$ & $S_{\rm MIR}$ & $\log M_{\rm tot}$ & $\log N$ \\
       & $\times$10$^{6}$ cm$^{-6}$\,pc & mJy & s$^{-1}$ & mJy & M$_{\odot}$ & O-stars \\
\hline
 M1 &    2.51 &    0.58 &   52.39 &  21.76 &   5.71 &   3.21 \\
 M2 &    2.44 &    0.56 &   52.38 &  21.11 &   5.69 &   3.19 \\
 M3 &    1.63 &    0.37 &   52.20 &  14.14 &   5.52 &   3.02 \\
 M4 &    1.61 &    0.37 &   52.20 &  13.93 &   5.51 &   3.01 \\
 M5 &    1.28 &    0.29 &   52.10 &  11.10 &   5.41 &   2.92 \\
 M6 &    0.98 &    0.22 &   51.98 &   8.49 &   5.30 &   2.80 \\
Envelope & 0.32 & 29.69 & 54.10 & 1122.50 & 7.42 & 4.92 \\
\hline
\end{tabular}
\end{table*}

The Lyman continuum photon flux required to maintain ionization is given by
\begin{equation}
N_{\rm ph} ({\rm s}^{-1}) = 8.04\times10^{46} T_{\rm e}^{-0.85}U^{3}.
\end{equation}
\noindent
The excitation parameter, $U$, can be written as 
\begin{equation}
U = 4.533\left[\left(\frac{\nu}{{\rm GHz}}\right)^{0.1} \left(\frac{T_{\rm e}}{{\rm K}}\right)^{0.35} \left(\frac{S_{\nu}}{{\rm Jy}}\right) \left(\frac{D}{{\rm kpc}}\right)^{2}\right]^{1/3}
\end{equation}
\citep*{kurtz94}, where $D$ is the distance to the source. 
Using the radio flux
calculated above, we find that the number of ionizing photons in the compact
MIR sources is $2 \times 10^{52}$ s$^{-1}$ as listed in the fourth column of 
Table~\ref{table:emtab}. 

Only massive O- and B-type stars contribute to the Lyman continuum flux, 
hence we can make an estimate of their numbers within each source. By making 
assumptions about the age of the compact sources and the initial mass 
function (IMF), we 
can also derive stellar composition and mass of each source.
In order to do this we run a Starburst99 (\citealt{Leitherer99,Vazquez05})
model as an instantaneous starburst with a Kroupa IMF \citep{kroupa02} with low 
mass cut-off at 0.10 M$_{\odot}$ and high-mass cut-off at 100 $M_{\odot}$. 
Assuming age of one million years, we find that the compact MIR sources 
contain 1000-1500 O-type stars. The total mass in stars of each source, 
including the stars that do not contribute to the ionizing flux, is 
typically 3--5$\times$10$^{5}$ M$_{\odot}$. The total stellar mass (assuming 
an age of one Myr) and the number of O-stars in each source are listed in the 
last two columns of Table~\ref{table:emtab}. 

The numbers vary relatively little with the age of the model as long as the 
age is $\la 4$ Myr. As the age increases, the Lyman continuum photon flux
decreases, so more stars are required in order to produce the observed photon 
flux of $\approx10^{52.4}$ s$^{-1}$. If the average photon flux per OB star
is $10^{49}$ photons\,s$^{-1}$ (\citealt*{smith02,martins05}), the cluster 
cannot contain more than a few thousand OB stars, as derived above. But if 
the age is $> 4$--5 Myr, the cluster must contain more OB stars and hence be
more massive.

If we assume that all the ionizing flux eventually is emitted into the 
infrared, we can calculate an upper limit to 
the MIR continuum. For this, we use eqs.\ A2 and A3 from \citet{genzel82} 
which assume that 
the dust emission can be approximated by a grey body radiating at 
300 K. The predicted upper limits at 12.8$\umu$m, labeled $S_{\rm MIR}$ in
Table~\ref{table:emtab}, are consistent with our measured 3$\sigma$ upper limit
of 0.27 Jy.

The radio flux from the central regions of NGC 7582 consists of 
contributions from thermal free-free emission from the starburst, synchrotron 
radiation from supernova remnants and the AGN. The
radio flux we derive above is, by definition, due to thermal free-free 
emission and we can compare the predicted thermal radio emission with 
the observed flux to estimate the contribution from non-thermal 
processes. As existing radio measurements cannot cannot resolve the different 
MIR components, we compare with the estimated 
thermal radio flux for the whole envelope (on the last line of 
Table~\ref{table:emtab}). At frequencies of 8.6, 4.9 and 1.4 
GHz we derive free-free continuum fluxes of 33.6, 29.7 and 28.1 mJy, 
respectively. The measured radio fluxes at the same frequencies are 49.29, 69 
and 166 mJy (\citealt{ulvestad_wilson84,morganti99}). The contribution from
free-free emission related to the starburst is therefore $\approx$20, 40 and 
70\% in order of increasing frequency. The relative contribution from 
free-free processes increases with frequency because, as opposed to 
synchrotron emission, 
free-free processes produce almost flat radio spectra. 
We have assumed that there is little synchrotron contribution from supernovae.
If the age is small enough that not many supernovae have exploded yet, this is
a valid assumption. 
However, there may be contributions to synchrotron emission from supernovae 
in the envelope. The observed extended \mbox{[Fe\,{\sc ii}]} emission 
\citep{sosabrito01} may indicate that some supernovae have already occurred.
It is however possible that the compact sources are younger than the 
surrounding envelope. 


\section{Conclusions}

We have detected several compact regions of \mbox[Ne\,{\sc ii}]12.8$\umu$m 
emission in the circum nuclear starburst disk of NGC~7582. The compact MIR 
sources are embedded in a larger and more diffuse envelope 300-400 pc across, 
and do not have counterparts detected at continuum optical and NIR wavelengths.
If the sources are young star clusters, the compact MIR emission 
suggests they are embedded in their own dusty birth material. The MIR emission 
of young, embedded clusters is expected to decrease as the most massive stars 
explode as supernovae and gas and dust 
are expelled. That the sources are bright in the MIR, and with no counterparts 
detected in continuum optical or NIR, suggests they may be young. It can 
however not be excluded that the clusters could be older and obscured by 
a screen of dust related to the circum nuclear disk.

We estimate that the number of ionizing photons within each compact source is
1--2.5$\times$10$^{52}$ s$^{-1}$. By comparing with starburst models we find 
that the stellar mass of the two brighter sources M1 and M2 probably is  
$\sim$5$\times$10$^{5}$ M$_{\odot}$ and that they each contain roughly 
1.5$\times$10$^{3}$ O-type stars. The less bright sources M3--M6 probably have
smaller masses of 2.0--3.2$\times$10$^{5}$ M$_{\odot}$ and contain 
$\sim$1.0$\times$10$^{3}$ O-type stars. These properties are very similar to 
those of young, embedded star clusters 
(\citealt{Keto99,Gorjian01,johnson_kobulnicky03}). 

Our observation highlights the need for high-resolution MIR 
imaging and spectroscopy to study nuclear starbursts and individual
embedded star clusters in circum-AGN environments. 

\section*{Acknowledgments}

We are grateful to the VISIR Science Verification team,
to M.~Lacy, A.~Bik and S.~Larsen for 
helpful discussions, and to D.J.~McKay for the archival ATCA image.
The referee is acknowledged for an insightful report.

\bibliographystyle{mn2e}
\bibliography{/Users/wold/publications/tex/references}
\bsp
\label{lastpage}
\end{document}